\documentclass[conference]{sig-alt-full}

\usepackage{graphicx}
\usepackage[tight,normalsize,sf,SF]{subfigure}
\usepackage{algorithm}
\usepackage{algpseudocode}
\usepackage{xspace}
\newcommand{\etal}{{\em et al.}}
\newcommand{\ignore}[1]{}
\newcommand{\GMC}{GreedyMaxClique\xspace}
\newcommand{\kCore}{\textit{k}-Core\xspace}
\newcommand{\CP}{CP\xspace}
\newcommand{\figref}[1]{Fig.~\ref{#1}}
\newcommand{\algoref}[1]{Alg.~\ref{#1}}

\algrenewcommand\algorithmicrequire{\textbf{Input:}}
\algrenewcommand\algorithmicensure{\textbf{Output:}}
\algrenewcommand\algorithmicforall{\textbf{foreach}}

\begin{document}
%
\title{Near-Deterministic Inference of AS Relationships}

\numberofauthors{3}
\author{
\alignauthor
Yuval Shavitt\\
\affaddr{School of Electrical Engineering}\\
\affaddr{Tel-Aviv University, Israel}\\
\email{shavitt@eng.tau.ac.il}
\alignauthor
Eran Shir\\
\affaddr{School of Electrical Engineering}\\
\affaddr{Tel-Aviv University, Israel}\\
\email{shire@eng.tau.ac.il}
\alignauthor
Udi Weinsberg\\
\affaddr{School of Electrical Engineering}\\
\affaddr{Tel-Aviv University, Israel}\\
\email{udiw@eng.tau.ac.il}
}

\maketitle

\begin{abstract}
The discovery of Autonomous Systems (ASes) interconnections and the
inference of their commercial Type-of-Relationships (ToR) has been
extensively studied during the last few years. The main motivation
is to accurately calculate AS-level paths and to provide better
topological view of the Internet. An inherent problem in current
algorithms is their extensive use of heuristics. Such heuristics
incur unbounded errors which are spread over all inferred
relationships. We propose a near-deterministic algorithm for solving
the ToR inference problem. Our algorithm uses as input the Internet
core, which is a dense sub-graph of top-level ASes. We test several
methods for creating such a core and demonstrate the robustness of
the algorithm to the core's size and density, the inference period,
and errors in the core.

We evaluate our algorithm using AS-level paths collected from
RouteViews BGP paths and DIMES traceroute measurements. Our proposed
algorithm deterministically infers over 95\% of the approximately
58,000 AS topology links. The inference becomes stable when using a
week worth of data and as little as 20 ASes in the core. The
algorithm infers 2$\sim$3 times more peer-to-peer relationships in
edges discovered only by DIMES than in RouteViews edges, validating
the DIMES promise to discover periphery AS edges.
\end{abstract}

\ignore{
}

\section{Introduction}
\label{sec:introduction}

Today's Internet consists of thousands of networks administrated by
various Autonomous Systems (ASes). ASes are assigned with one or
more blocks of IP prefixes and communicate routing information to
each other using Border Gateway Protocol (BGP). Each AS uses a set
of local policies for selecting the best route for each reachable
prefix. Typically, these policies are based on the
Type-of-Relationship (ToR) that exists between ASes and on a
shortest path criteria. In order to calculate the paths between
ASes, one needs to obtain the ToR between all neighboring ASes.
Since ToRs are regarded as proprietary information, deducing them is
an important yet difficult problem.

Typically~\cite{inter}, there are three major commercial
relationships between neighboring ASes: customer-to-provider (c2p),
peer-to-peer (p2p), and sibling-to-sibling (s2s). In the c2p
category, a customer AS pays a provider AS (usually larger than the
customer) for traffic that is sent between the two. In the p2p
category, two ASes freely exchange traffic between themselves and
their customers, but do not exchange traffic from or to their
providers or other peers. In s2s, two ASes administratively belong
to the same organization and freely exchange traffic between their
providers, customers, peers, or other siblings.

Gao~\cite{gao01} was the first to study the AS relationships
inference problem and deduced that every BGP path must comply with
the following hierarchical pattern: an uphill segment of zero or
more c2p or s2s links, followed by zero or one p2p links, followed
by a downhill segment of zero or more p2c or s2s links. Paths with
this hierarchical structure are called \emph{valley-free} or valid.
Paths that do not follow this hierarchical structure are called
invalid and may result from BGP misconfigurations or from BGP
policies that are more complex and do not distinctly fall into the
above classification. Most work in this field
(section~\ref{sec:related}) follows the valley free routing
principle.

Current relationships inference algorithms attempt to solve the
ToR problem either by using heuristic assumptions or by optimizing
some aspects of the ToR assignments. Optimization is usually
achieved by minimizing the number of paths that violate the valley
free routing property~\cite{subramanian02characterizing} while not
allowing cycles to be created~\cite{ator-2007,KMT-2006} in the
resulting directed relationships graph.

Using heuristic assumptions throughout the relationships inference
process causes the erroneous ToRs to be spread over all
interconnecting ASes links. The optimization models fail to capture
the true Internet hierarchy~\cite{dimitropoulos-2005-3503} and have
a relatively low p2p inference accuracy~\cite{XiaGao04}. The result
is that both solutions fail to provide an insight, or a bound on the
inference errors.

Typically, AS relationships are not published by AS operators, hence
the validation of such results is done either by sending queries to
the operators of a small subset of ASes~\cite{dimitropoulos-2006-37}
or by comparing the results to partial information that is available
on the Internet~\cite{XiaGao04}. Although these methods can give a
good approximation on the correctness of the results, one cannot
assume a bounded mistake.

This paper aims to improve on existing methods by providing a
near-deterministic inference scheme for solving the ToR problem. The
input for our algorithm is the Internet \emph{Core}, a sub-graph
that consists of the globally top-level providers of the Internet
graph and their interconnecting edges with their already inferred
relationship types. Theoretically, given an accurate core with no
relationships errors, the algorithm \emph{deterministically} infers
most of the remaining AS relationships using the AS-level paths
relative to this core, without incurring additional inference
errors. In real-world scenarios, where the core and AS-level paths
can contain errors (due to misconfigurations or measurements
mistakes), the algorithm introduces minimal inference mistakes. The
core can be approximated in several ways, as described in
section~\ref{subsec:core-graph}, or extracted from public databases.
We show that our algorithm has relaxed requirements from the core,
and proves to be robust under changes in its definition, size and
density. Since the top-level ASes are a small and stable group,
accurately revealing the core members and their mutual types of
relationships is fairly easy. For the remaining set of relationships
that cannot be inferred deterministically, a heuristic inference
method is deployed. Since this group is relatively small, it is
possible to provide a strict bound on the inference error. In order
to increase the number of vantage points from which we see the
Internet, we use both RouteViews (RV)~\cite{routeviews} BGP data and
DIMES~\cite{1096546} AS-level traceroutes. We expect that over time,
the group of non-deterministic inferred relationships will even
further decrease.

\ignore{ REMOVED - is this important?? We evaluate the proposed ToR
inference methods in several aspects. We first investigate the
effect that the core selection (i.e., its vertices and edges) has
over the number of edges we can classify without using heuristics.
We then look at the effect of the amount of aggregated data we use
in order to classify the Internet AS-level graph as an attempt to
detect changes that occur over time in the relationships between
ASes. We also examine the accuracy of the algorithm and compare it
to other previous works, and provide additional analysis of edges
that are seen using DIMES data. }

The remaining of this paper is organized as follows.
Section~\ref{sec:related} provides several related works concerning
AS relationships inference. Section~\ref{sec:inferrence} provides a
detailed description of our deterministic inference algorithm and
discusses the methods used to infer on the remaining unclassified
edges. Section~\ref{section:exp-results} provides a detailed
evaluation of the proposed algorithms and
Section~\ref{sec:conclusion} concludes the paper and discusses
future work.

\section{Related Work}
\label{sec:related} As mentioned above, Gao's pioneering work
\cite{gao01} was the first to study the AS relationships inference
problem. Gao proposed an inference heuristic that identified top
providers and peering links based on AS size, which is proportional
to its degree (the number of immediate neighbors of a vertex), and
the valley-free nature of routing paths. Gao used this heuristic to
infer relationship between ASes in the Internet by traversing
advertised BGP routes, locally identifying the top provider for each
path, and classifying edges (i.e., inferring the relationships
represented by the edges) as going uphill to the top provider and
downhill afterwards. Xia and Gao~\cite{XiaGao04} later proposed to
use partially available information regarding AS relationships in
order infer the unknown relations. It is not clear that this
information can be obtained and validated periodically, unlike our
suggestion to use the almost constant relationships in the Internet
core. Our sole reliance on the core produces simpler inference rules
that are less prone to inference errors.

Following Gao's work, Subramanian \etal\
\cite{subramanian02characterizing} formally defined the
Type-of-Relation (\emph{ToR}) maximization problem that attempts to
maximize the number of valid (valley-free) routing paths for a given
AS graph. Their approach takes as input the BGP tables collected at
different vantage points and computes a rank for every AS. This rank
is a measure of how close to the graph core an AS lies (equivalent
to vertex coreness \cite{alvarezhamelin-2005}), and is heuristically
used to infer AS relationships by comparing ranks of adjacent ASes.
If the ranks are similar, the algorithm classifies the link as p2p,
otherwise it is classified as c2p or p2c.

Battista \etal~\cite{battista02computing} showed that the decision
version of the ToR problem (\emph{ToR-D}) is an NP-complete problem
in the general case. Motivated by the hardness of the general
problem, they proposed approximation algorithms and reduced the
ToR-D problem to a 2SAT formula by mapping any two adjacent edges in
all input AS-level routing paths into a clause with two literals,
while adding heuristics based inference.

Dimitropoulos \etal~\cite{dimitropoulos-2005-3503} addressed a
problem in current ToR algorithms. They showed that although ToR
algorithms produce a directed Internet graph with a very small
number of invalid paths, the resulting AS relationships are far from
reality. This led them to the conclusion that simply trying to
maximize the number of valid paths (namely improving the result of
the ToR algorithms) does not produce realistic results. Later in
\cite{dimitropoulos-2006-37} they showed that ToR has no means to
deterministically select the most realistic solution when facing
multiple possible solutions. In order to solve this problem, the
authors suggested a new objective function by adding a notion of "AS
importance", which is the AS degree "gradient" in the original
undirected Internet graph. The modified ToR algorithm directs the
edges from low importance AS to a higher one. The authors showed
that although they have high success rate in p2c inference (96.5\%)
and in s2s inference (90.3\%), the p2p inference success rate
(82.8\%) is relatively low. Moreover, the authors surveyed some ASes
operators and mention that for some of them, the BGP tables, which
are the source for AS-level routing paths for most works in this
research field, miss up to 86.2\% of the true relationships between
adjacent ASes, most of which are of p2p type.

These observations match the evaluation work done
in~\cite{XiaGao04}, and highly motivate our work, driving us not
only to seek an algorithm that better captures the true AS
relationships in the Internet while reducing the usage of heuristics
for inference, but also to add a different, complementary data
source for routing paths, that has the ability to capture much of
the missing links.

\ignore{
 Cohen and Raz \cite{ator-2007} follow previous works
\cite{gao00stable,gao01inherently} and describe an algorithm that
attempts to minimize the number of invalid paths, but in addition,
captures the true hierarchal structure of the real Internet.
Following the fact that the real Internet graph cannot contain
cycles, they defined the Acyclic Type of Relationship (\emph{AToR})
problem as an attempt to maximize the number of valid paths from a
given set of routing paths, while keeping the directed graph
acyclic. A parallel work that provided a very similar definition to
the AToR problem, but was very focused on the theoretical aspects of
the problem was published by Kosub \etal\~\cite{KMT-2006}. It was
later applied to the Internet~\cite{TUM-I0709}, showing that the
acyclicity assumptions are valid on the Internet graph. }

\section{AS Relationships Inference}
\label{sec:inferrence} In this section we describe our ToR inference
algorithm in details. We start with explaining the deterministic
algorithm, and proceed with the heuristics we employ for edges that
the deterministic algorithm fails to classify.

\subsection{Deterministic Classification}
\label{subsec:deterministic-algorithm}Our deterministic algorithm
receives as input two undirected AS-level graphs and a set of
AS-level routing paths, denoted by $S$. The first graph, denoted by
$G(V_{G},E_{G})$, contains the set of vertices that represent all
ASes, and the interconnecting edges that need to be classified. The
second graph, denoted by $Core(V_{C},E_{C})$, holds the vertices and
interconnecting edges that represent the core of $G$, and is assumed
to contain all the top-level ASes.

Prior to starting the relationships inference algorithm, we infer
s2s relationships, since ignoring these relationships might cause
proliferation of erroneous inference~\cite{dimitropoulos-2006-37}.
We use s2s data collected from~\cite{caida-as-rank}. These s2s
classifications are obtained from IRR databases, namely RIPE, ARIN
and APNIC. Although these databases are not always up-to-date, they
are reasonably steady and accurate for the s2s inference. Once
classified, the s2s edges are removed from the edges set $E_{G}$,
and the two adjacent vertices are united to form a single vertex
that inherits the connectivity of both.

Following the assumption that the input core consists of all the
global top-level ASes and using the valley-free model of Internet
routing, the algorithm classifies most of the edges in
$G(V_{G},E_{G})$ without using heuristic assumptions:

\textbf{Phase 1}. All paths that \emph{pass through the core} are
split into a segment of zero or more uphill c2p edges towards the
core, at most one p2p edge in the core and a downhill segment of
zero or more p2c edges from the core. The algorithm, shown in
\algoref{alg:phase1}, traverses only paths that pass through the
core. It starts with the uphill segment of the path, classifying
each edge as c2p, until reaching the core. Once reached the core,
the uphill segment finishes and the core segment starts. Inside the
core the algorithm classifies edges that are not already classified
as p2p (the default type for core edges). The downhill segment
starts with an AS that does not belong to the core, and is traversed
until the end of the path. Invalid paths are detected when an edge
is directed towards the core (uphill) during the downhill segment.
Each path that is classified in this phase is removed from the set
of paths $S$. Note that the algorithm does not use direct inference
but a voting technique (see section~\ref{subsec:determining-classification})
in order to resolve ambiguities resulted from incorrect paths.
\begin{figure}[t]
\centering \subfigure[Phase 1]{
    \label{fig:traverse_the_core}
    \includegraphics[width=.20\textwidth]{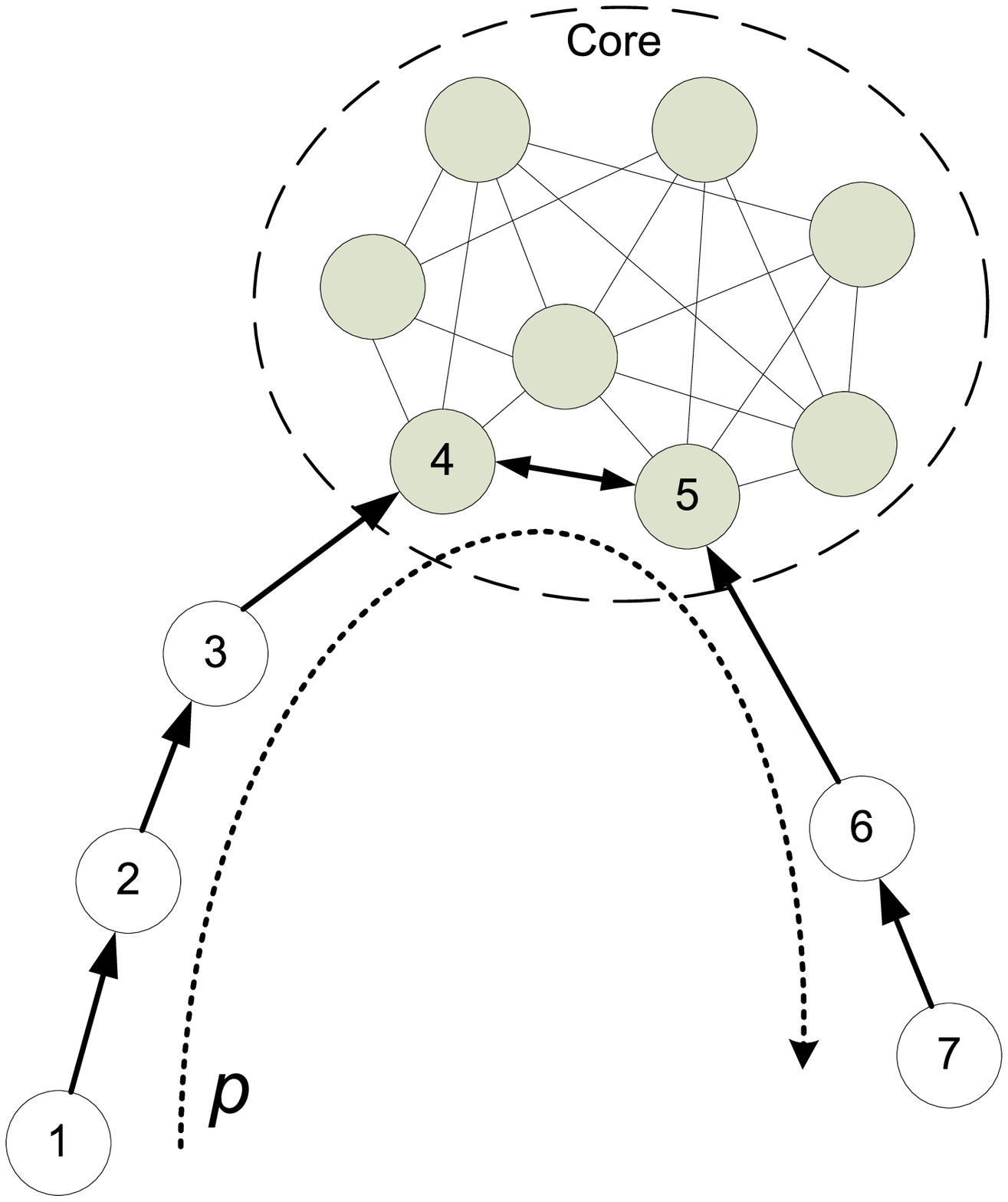}}
    \hspace{.3in}
    \subfigure[Phase 2]{
    \label{fig:overlap_traverse_the_core}
    \includegraphics[width=.20\textwidth]{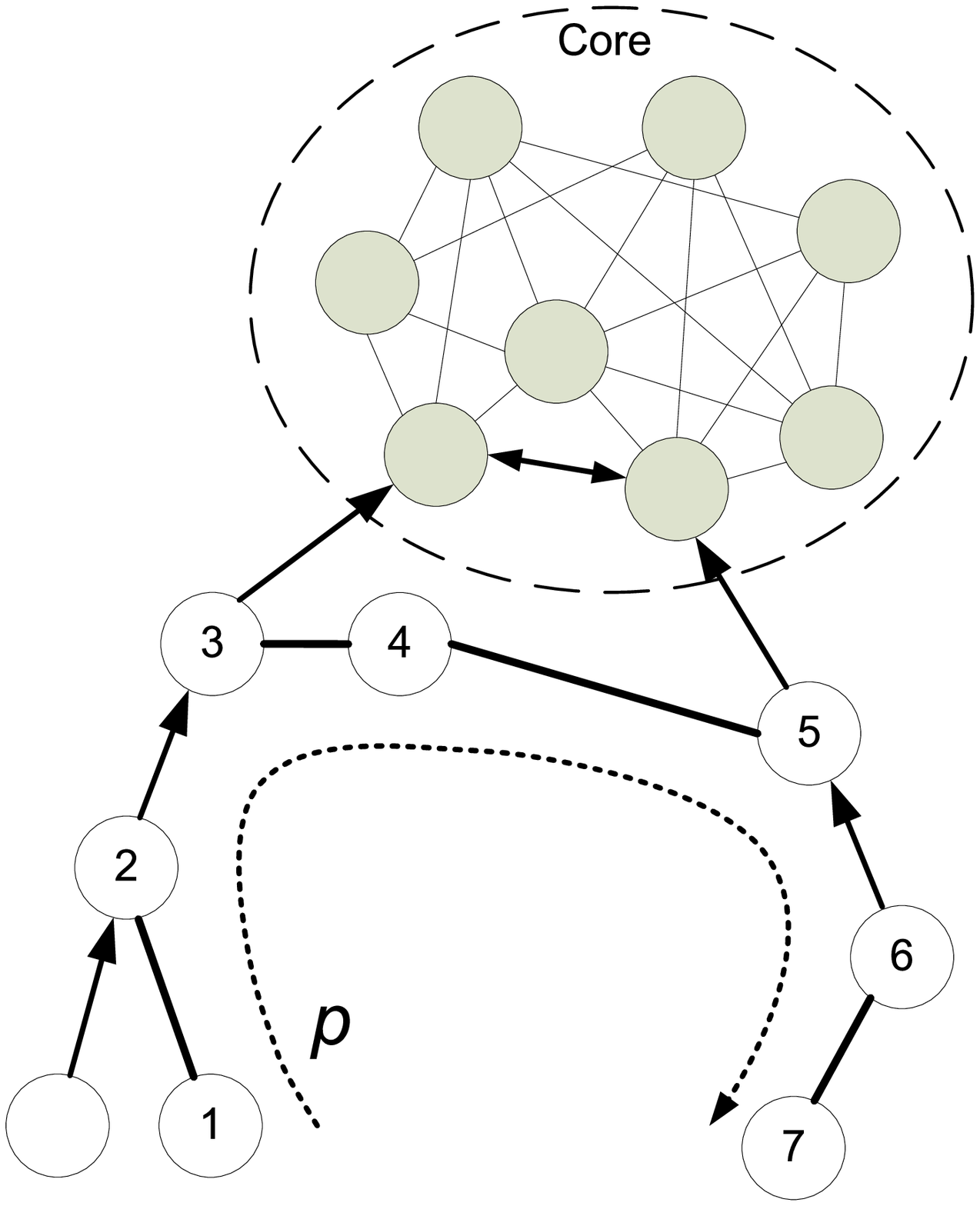}}
    \caption{Deterministic ToR inference algorithm}
    \label{fig:algorithm}
\end{figure}

An example for edges that are classified during this phase is
illustrated in \figref{fig:traverse_the_core}. Path \emph{P}
traverses the core and consists of seven AS hops, numbered 1 to 7.
The segment between hop 1 and hop 4 is identified as an uphill
segment, resulting in the classification of the edges (1,2),
(2,3), and (3,4) as c2p (illustrated as an arrow pointed from the
customer to the provider). ASes 4 and 5 are inside the core,
therefore are already classified (by default to p2p). The segment
starting in AS 5 and ending in AS 7 is considered as a downhill
segment, resulting in the classification of the edges (5,6) and
(6,7) as p2c.
\begin{algorithm}
\centering \scriptsize
\begin{algorithmic}[1]
\Require Graphs $G(V_{G},E_{G})$, $Core(V_{C},E_{C})\subset G$,
Paths set $S$ \Ensure Edges $E_{G}$ with votes for relationship
types
\ForAll{$path \in S$}
\If{$\exists e \in path\mid e\in
E_{C}~\textbf{or}~\exists v \in path\mid v\in V_{C}$}
  \State $upHill \gets TRUE$
  \State $downHill \gets FALSE$
  \State $inCore \gets FALSE$

  \ForAll{$edge \in Path$}
    \State $AS1 \gets edge.firstAS$
    \State $AS2 \gets edge.secondAS$

    \If {$edge \in E_{C}$}
        \State $upHill \gets FALSE$
        \State $inCore \gets TRUE$
    \ElsIf {$AS1 \in V_{C} \mathbf{~and~} AS2 \not \in V_{C}$}
        \State $upHill \gets FALSE$
        \State $inCore \gets FALSE$
        \State $downHill \gets TRUE$
    \ElsIf {$downHill \mathbf{~and~} AS2 \in V_{C}$}
        \State $voteForInvalid(edge)$
    \EndIf
    \If {$upHill$}
        \State $voteForCustomerToProvider(edge)$
    \ElsIf {$inCore~\textbf{and}~notClassified(edge)$}
        \State $voteForPeerToPeer(edge)$
    \Else
        \State $voteForProviderToCustomer(edge)$
    \EndIf
  \EndFor
  \State $S \gets S \setminus path$
  \EndIf
  \EndFor
\caption{Phase 1 of ToR Inference Algorithm}
\end{algorithmic}
\label{alg:phase1}
\end{algorithm}

Since the remaining paths in $S$ do not traverse the core, they do
not provide us with a direct method for classification. However,
amongst these, there are paths that partly overlap other paths
that traverse the core. Meaning that some of the remaining paths
already contain edges that were classified as either c2p or p2c in
the first phase of the algorithm. We use these edges for the
second phase of the algorithm:

\textbf{Phase 2}. For a given path, edges that precede a c2p edge
must reside in an uphill segment, and be of type c2p. Edges that
follow a p2c edge must be in a downhill segment, and be of type
p2c. The algorithm, listed in \algoref{alg:phase2}, traverses one
path at a time, and looks for an already inferred c2p or p2c
edges. If a c2p edge is detected, all unclassified edges in the
path before this edge, temporarily stored in the \emph{suspectC2P}
list, are classified as c2p. If a p2c edge is detected, all
unclassified edges in the path after this edge, temporarily stored
in the \emph{suspectP2C} list, are classified as p2c.

Since this phase uses classified edges in order to classify
unclassified edges, it is repeated for all paths in $S$ that still
have unclassified edges, until there are no more edges that can be
classified using this method.
\begin{algorithm}
\centering \scriptsize
\begin{algorithmic}[1]
\Require Graph $G(V_{G},E_{G})$, Remaining set of paths $S$
\Ensure Edges $E_{G}$ with votes for relationship types
\ForAll{$path \in S$}
  \State $suspectC2P \leftarrow \emptyset$
  \State $suspectP2C \leftarrow \emptyset$
  \State $passedP2C \leftarrow FALSE$
  \ForAll{$edge \in Path$}
    \If {$maxVotesC2P(edge)\mathbf{~and~}suspectC2P\neq\emptyset$}
        \ForAll{$e \in suspectC2P$}
            \State $voteForCustomerToProvider(edge)$
        \EndFor
        \State $suspectC2P \leftarrow \emptyset$
    \ElsIf {$maxVotesP2C(edge)$}
        \State $suspectC2P \leftarrow \emptyset$
        \State $passedP2C \leftarrow TRUE$
    \EndIf
    \If {$noClassificationVotes(edge)$}
        \If {$passedP2C=FALSE$}
            \State $suspectC2P \leftarrow suspectC2P\cup edge$
        \Else
            \State $suspectP2C \leftarrow suspectP2C\cup edge$
        \EndIf
    \EndIf
  \EndFor
  \If {$suspectP2C\not\equiv\emptyset$}
    \ForAll{$e \in suspectP2C$}
        \State $voteForProviderToCustomer(edge)$
    \EndFor
  \EndIf
\EndFor \caption{Phase 2 of Classification Algorithm}
\end{algorithmic}
\label{alg:phase2}
\end{algorithm}

\figref{fig:overlap_traverse_the_core} illustrates an example for a
path that contains edges that will be classified during Phase 2 of
the algorithm. Path \emph{P} does not traverse the core but contain
edges that are already classified. The edge (1,2) precedes the edge
(2,3) which is classified as c2p, therefore is classified as c2p.
The edge (6,7) follows a p2c edge, therefore is classified as p2c.
The edges (3,4) and (4,5) cannot be classified, since we are unable
to determine which AS is the top-level provider and whether the
edges (3,4) and (4,5) represent a p2p relationship.

\subsection{Assigning Type-of-Relationship to Edges}
\label{subsec:determining-classification} The data we use might be
"noisy" and reflect transient routing effects or changes in the
commercial relationships between ASes, especially when performing
relationships inference over a long time frame. To avoid incorrect
inferences resulting from these effects, we use voting
technique~\cite{gao01} instead of direct relationship inference.
Meaning, that the above methods vote for the ToR of each traversed
edge. Once the algorithm is finished, we count the votes and assign
each edge with the type that received a relative votes count that
passes a given threshold. \ignore{The pseudo-code of this vote
counting technique is provided in \algoref{alg:voting}. The
algorithm simply calculates the relative vote count for each type of
relationship and returns the one that exceeds the provided
threshold.}

Note that the voting technique is not used as a heuristic inference
method, but rather as a method to avoid inference bias caused by
incorrect or transient paths. The deterministic algorithm does
not attempt to classify edges that have a vote count that does not
pass the given threshold, but rather it employs the non-deterministic methods
discussed in section~\ref{sec:unclassified-algorithm}. Setting the exact value
of the threshold, in order to achieve these goals is further discussed in
section~\ref{subsec:vote-threshold}.
\ignore{
\begin{algorithm}
\centering \scriptsize
\begin{algorithmic}[1]
\label{alg:voting}
\Require Edge $e$, $threshold$ \Ensure Type-of-Relationship for
edge $e$ \State $p2c \leftarrow 0$
  \State $c2p \leftarrow 0$
  \State $p2p \leftarrow 0$
  \State $sum \leftarrow (e.p2p+e.p2c+e.c2p)$
  \If {$sum \neq 0$}
    \State $p2c \leftarrow e.p2c/sum$
    \State $c2p \leftarrow e.c2p/sum$
    \State $p2p \leftarrow e.s2s/sum$
    \If {$p2c\geq threshold$}
        \State $\textbf{Return} provider-to-customer$
    \ElsIf {$c2p\geq threshold$}
        \State $\textbf{Return} customer-to-provider$
    \ElsIf {$p2p\geq threshold$}
        \State $\textbf{Return} peer-to-peer$
    \Else
        \State $\textbf{Return} heuristicsInference(e)$
    \EndIf
  \Else
    \State $\textbf{Return} unclassified$
  \EndIf
\caption{Relation inference for a given edge}
\end{algorithmic}
\end{algorithm}
}

\ignore{ In the following section we identify the edges that the
non-heuristic algorithm fails to classify, and discuss methods that
help us classify the remaining edges.

The experimental results discussed in section
\ref{section:exp-results} show that less than TBD\% of the edges
remain unclassified before employing additional non-deterministic
classification techniques, and only TBD\% remains after.}

\subsection{Non-Deterministic Inference of Remaining Relationships}
\label{sec:unclassified-algorithm} 
\begin{figure}[tbh]
\centering \subfigure[Peer-to-Peer]{
    \label{fig:bridge}
    \includegraphics[width=.21\textwidth]{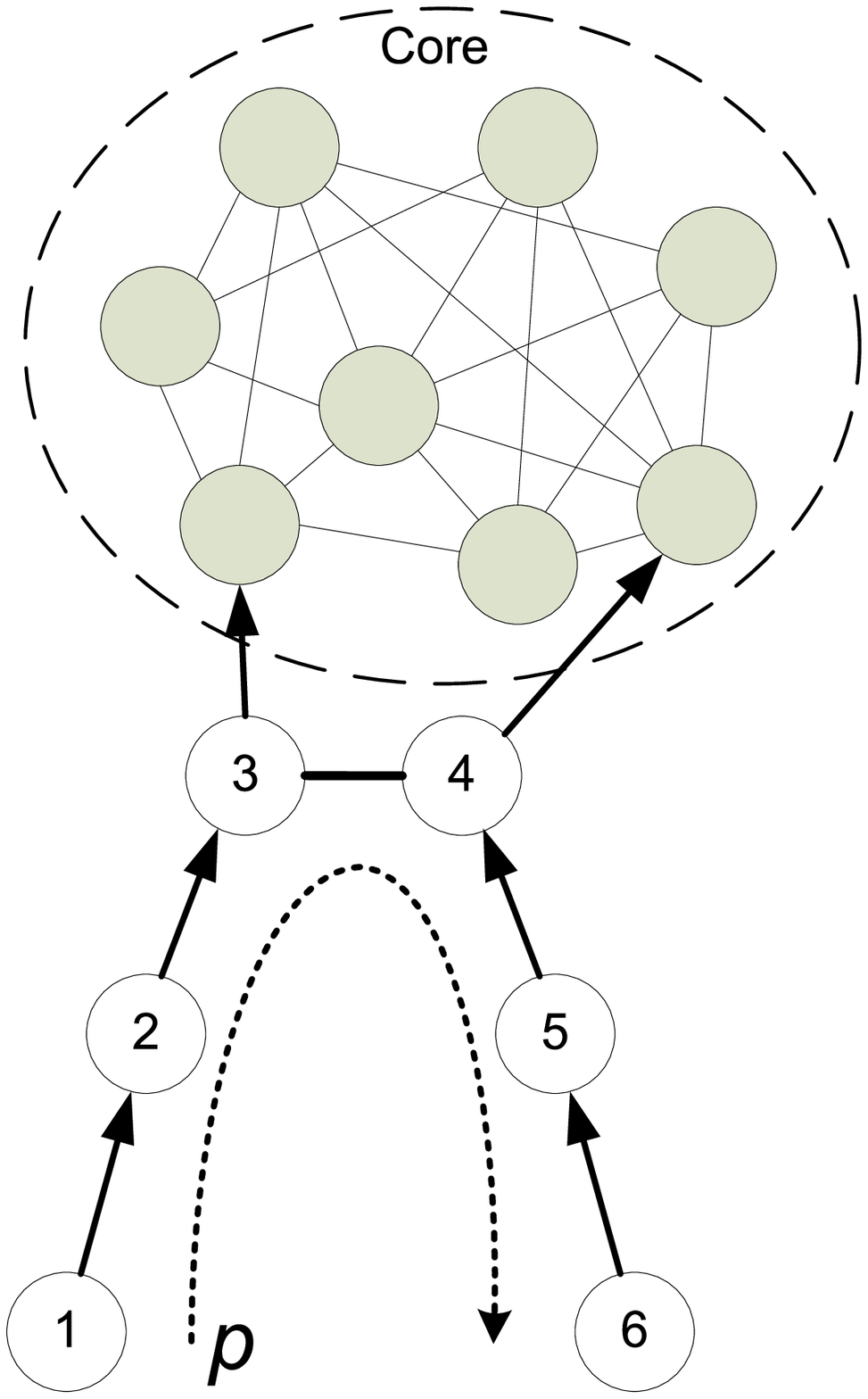}}
    \hspace{.1in}
    \subfigure[Non-Valley-Free]{
    \label{fig:valley}
    \includegraphics[width=.21\textwidth]{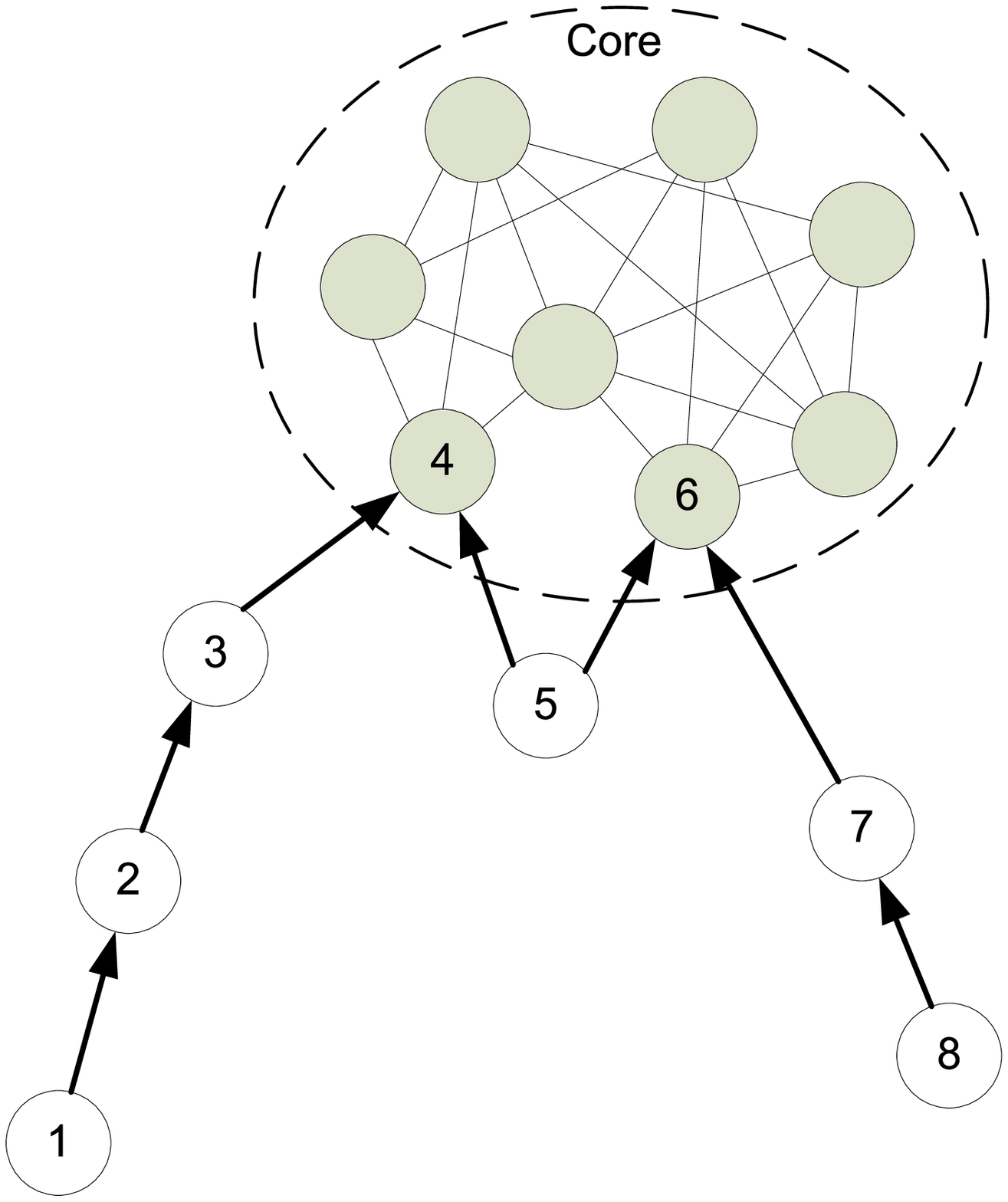}}
    \caption{Non-Deterministic ToR inferences}
    \label{fig:non-deterministic}
\end{figure}

The deterministic algorithm fails
to classify several types of edges. The first type are edges that
appear in paths that do not traverse the core, and reside between a
c2p edge and a p2c edge (see \figref{fig:bridge}). This can be a
result of a path that does not traverse the core and has no
overlapping edges with other paths, or overlaps other paths in edges
that are close to the beginning or end of the path. Alternatively,
the path may have a p2p relationship between its two top-level
vertices.

In order to infer relationships related to these edges we use the
following assumption (which is backed from observations, as we show
in section~\ref{subsec:exp-unclassified-edges}): a c2p or p2c edges
should participate in, at least, one path that pass through the
core. For this not to happen the following should occur: 1) a client
will rarely route through a certain edge to its provider and thus
may not expose this link in a DIMES measurement path that passes
through the core, and 2) at the same time the paths through the core
that contain this edge will be filtered by BGP in the direction of
the speakers sampled by RouteViews. Thus, we assume that most c2p
and p2c edges are already classified by our deterministic algorithm.
Following this assumption we can infer that in paths that do not
pass through the core, and have a \emph{single} remaining
unclassified edge (which must reside between a c2p and a p2c edges),
this edge should be classified as a p2p edge (this is illustrated as
the edge between ASes 3 and 4 in \figref{fig:bridge}). In case there
is more than a single edge between the c2p and the p2c (as
illustrated in \figref{fig:overlap_traverse_the_core}), we leave the
edges unclassified, since we cannot determine which of the vertices
is the provider.

The second type of unclassified edges are the ones that have a
similar number of votes for two or more types of relationships. This
can be the result of changes in the commercial relationship between
adjacent ASes over the measurements period, or due to more complex
peering agreements that can cause the same edge to behave
differently as seen from different view points in the
Internet~\cite{dimitropoulos-2006-37}.

To resolve these ambiguities, we use heuristic-based methods
suggested by other works. Although we chose to use the AS degree in
the graph~\cite{gao01}, and the k-shell
index~\cite{jellyfish,carmi-2006}, any other method can be employed.
Analysis of this inference technique is further discussed in the
experimental results in section~\ref{section:exp-results}.

The third type of unclassified edges are edges that appear in
non-valley-free paths (\figref{fig:valley}), possibly the result of
valid paths that pass a malformed core, or invalid paths that pass
an accurate core. Since these invalid paths occur in only a small
fraction of paths (less than 1\% on average from the investigated
paths per week), we leave the classification of these "valley edges"
to future work.

\subsection{Core Graph Construction} \label{subsec:core-graph}
Motivated by the need to capture the true global hierarchal
structure of the Internet we looked for an accurate global
decomposition of the Internet AS-level graph. There have been
several attempts to characterize the core of the Internet AS graph
\cite{subramanian02characterizing,jellyfish,carmi-2006,govindan97analysis,ge01hierarchical}.
We use three core construction methods, that result in cores that
vary in size and density. We analyze the effect that each core has
on the classification algorithm.

Tauro \etal~\cite{jellyfish} proposed the \emph{Jellyfish}
conceptual model in which they identified a topological center and
classified vertices into layers with respect to the center. The
authors defined core as a clique of high-degree vertices, and
constructed it by sorting the vertices in non-increasing degree
order. The first vertex in the core is the one with the highest
degree. Then, they examine each vertex in that order; a vertex is
added to the core only if it forms a clique with the vertices
already in the core. The resulting core is a clique but not
necessarily the maximal clique of the graph. We refer to this core
as \emph{\GMC}.

Carmi \etal~\cite{carmi-2006} indicated that using the popular
vertex's degree (which was encouraged by the finding of the
Internet's power-law distribution \cite{power-law}) as an
indicator of the vertex's importance can be misleading. The
authors presented the new \emph{Medusa} model, that uses a
$k$-pruning algorithm to decompose the Internet AS graph and
extract a nucleus (the $K_{max}$-Core) which is a very well
connected globally distributed subgraph. Note that this algorithm
extracts a core by looking at the entire graph, unlike \GMC that
takes a local approach. The properties listed for this model are
useful for AS relationship inference, mainly due to the finding
that the nucleus plays a critical role in BGP routing, since its
vertices lie in a large fraction of the paths that connect
different ASes. We refer to this core as \emph{k}-Core.

The last core we use is constructed from most of the ASes and
interconnecting edges that exhibit p2p relationship under the
inference method in~\cite{dimitropoulos-2006-37}. We use the
Automated AS ranking provided by CAIDA \cite{caida-as-rank} and
constructed a graph that contains all the edges classified as p2p
\ignore{(tagged with 0 in the files downloaded from
$http://as-rank.caida.org$)} and their adjacent AS vertices. We then
selected the largest connected component that contains some of the
largest tier-1 ASes, namely AS701 ($UUNET$) and AS7018
($AT\!\&\!T$). We refer to this core as \emph{\CP} (CAIDA Peers).

The three core types vary in size and density as an attempt to
capture different inference behaviors. Using a small, dense core
reduces the probability that a non-top-level AS is wrongfully
considered as a top-level AS for all paths that pass through it,
thus causing incorrect inferences. However, a small core might miss
top-level ASes, thus cause non-valley-free paths. On the other hand,
when using a large core, a trace might have several hops in the
core. In this case we follow~\cite{subramanian02characterizing} and
assume that two ASes may have an "indirect peering" relation,
meaning they have p2p relationship through an intermediate AS, such
as an exchange point. Traces with more than three hops in the core
are considered invalid.

\section{Experimental Results}
\label{section:exp-results} In this section we evaluate the
deterministic algorithm and the additional heuristics inferences
using data from the first five weeks of 2007. We evaluate its
accuracy by comparing the results to the classification algorithm
proposed in \cite{dimitropoulos-2006-37}, referred to as $CAIDA$. We
start by discussing the data sources, and the three different type
of cores we use as inputs to the algorithm. We then analyze the
effect of the core size (number of vertices) and density (the number
of edges divided by the full clique size) on the algorithm, and
check the transient effects caused by aggregating data for changing
time frames. Finally we check the sensitivity of the algorithm to
increasing mistake in the core.

\subsection{Data Sources}
\label{subsec:data-sources} For this work we combined data from the
RouteViews (RV)~\cite{routeviews} and DIMES~\cite{1096546} projects
to maximize the size of our AS topology. We used BGP paths collected
by the RV project, similar to most other previous
work~\cite{dimitropoulos-2005-3503,gao01inherently,gao00stable,jellyfish,power-law,subramanian02characterizing,1064257}.
We created weekly batches of AS-level paths by downloading one RV
file that was generated daily at 20:00, and merged all 7 files,
making sure that each path appears exactly once. We parse RV's files
and use only AS paths marked as "valid". \ignore{TBD-REMOVE THIS??
Some of the paths advertised in BGP contain a repeating AS. This is
caused due to routers that prepend their AS number multiple times to
make the route longer, thus discourage the route being selected as
the best route~\cite{1064257}. We process these paths and drop
repeating ASes.}

Since BGP paths miss many of the actual links (primarily of type
p2p) caused by non-advertised links in BGP
\cite{dimitropoulos-2006-37,ator-2007} we use additional data from
DIMES. DIMES is a large-scale distributed measurements effort that
measures and tracks the evolution of the Internet from hundreds of
different view-points, in an attempt to overcome the "law of
diminishing returns"~\cite{505204}. DIMES daily collects over 2
million traceroute and ping measurements targeted at a set of over 5
million IP addresses, which are spread over all the allocated IP
prefixes.

In order to create AS-level paths from the IP-level traceroutes
provided by DIMES agents, we preform AS resolution for each hop in
all paths. Although IP-to-AS mapping can be a difficult task~\cite{mao04scalable}
we take a strait-forward resolution approach, and then filter out paths
that exhibit various animalities.
AS resolution is done by first performing longest-prefix-matching
against BGP tables obtained from RV archive.
This resolves approximately 98\% of the IP addresses. For the
remaining 2\%, we query against two WhoIs databases, namely RIPE and
RADB, that resolves additional 1.5\% of the IP addresses. The
remaining 0.5\% unresolved IP addresses are discarded and do not
participate in the inference algorithm.

The raw DIMES data was filtered in order to reduce inference
mistakes and inclusion of false links. We filtered for some
measurements artifacts by only including edges which were seen from
at least two agents. In addition we trimmed all traces that exhibit
known traceroute problems~\cite{paris}, namely routing loops and
destination impersonation, keeping only the section of the path
preceding the identified problem.

\ignore{Using AS paths gathered from DIMES gives us the ability to
identify and classify edges that might not be advertised in BGP,
thus enabling us to obtain a complete AS-level Internet graph, as
seen by numerous vantage points, located at various types and sizes
of ASes.

Using this data, we have the ability to construct the inputs
required by our algorithm: the path set, $S$, consists of all paths
from both data sources gathered in the examined time frame and
passed the filtering rules; the AS Internet graph, $G$, is the
directed graph consists of both directions of every edge that is
contained in some RouteViews or DIMES path during the examined time
frame; the core graph $Core$ construction is discussed in the
following section.}

Table~\ref{table:edges-count} shows the number of ASes and
interconnecting links gathered during the first five weeks of
2007, obtained by using both RouteViews and DIMES. On average, the
data set consists of over 24,000 AS vertices and approximately
58,000 links (undirected edges). Approximately 44\% of the edges
exist only in RV paths, about 12\% exist only in the filtered
DIMES paths and the remaining 44\% of the edges exist in both RV
and DIMES. In section~\ref{subsec:exp-unclassified-edges} we
analyze the edges seen only by DIMES in order to understand the
type of these additional links.
\begin{table}[htb]
\caption{ASes and links collected by Dimes and RouteViews during
the first five weeks of 2007} \label{table:edges-count}
 \centering \scriptsize
\begin{tabular}{|c|p{0.8cm}|p{0.8cm}|p{1.1cm}|p{1.1cm}|p{1.2cm}|}
\hline
Week & ASes & Links & RV links & DIMES links & RV\&DIMES links\\
\hline
1 & 24391 & 57875 & 24282 & 6964 & 26629 \\
2 & 24451 & 57920 & 24313 & 6986 & 26621  \\
3 & 24492 & 57921 & 24609 & 6630 & 26682  \\
4 & 24581 & 59058 & 24913 & 7288 & 26857  \\
5 & 24716 & 59779 & 25528 & 7331 & 26920  \\
\hline
\end{tabular}
\end{table}

On a weekly average, we filtered approximately 5,100 DIMES edges
that were measured only once, which is over 15\% of the edges
measured by DIMES. Around half of these edges appear in RouteViews,
providing a testimony to our conservatism.

\subsection{Voting Threshold} \label{subsec:vote-threshold}
\begin{figure}[htb]
\centering
\includegraphics[width=.35\textwidth]{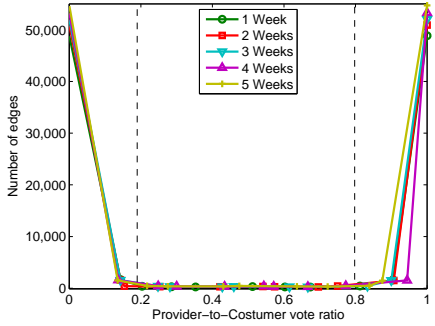}
\caption{Types-of-Relations voting distribution} \label{fig:voting}
\end{figure}

In order to validate the usage of the voting technique described in
section~\ref{subsec:determining-classification} and set a proper
threshold value, we tested the distribution of votes to inference
types. For each edge we calculated the ratio of p2c votes.
\figref{fig:voting} shows the number of edges for each p2c ratio.
Clearly, the vast majority of the edges are uniquely classified as
either p2c or c2p. This remains true when running the algorithm on
longer time frames.

Looking at the data backing up this graph, we see that on average
over 94\% of the edges have votes for exactly one relationship type,
and almost 99\% of the edges have over 80\% of the votes casted for
a single relationship type, which provides a very high level of
confidence for this selected type. Thus, a threshold value of 0.8
covers almost 99\% of the edges, and leaves approximately 1\% of the
edges to be classified using heuristic methods, or remain
unclassified.

\subsection{Sensitivity Analysis}
\label{subsec:sensitivity-core-size} Since the construction of the
core graph is a major building block for the algorithm, we evaluate
the effect that the core has on the inference process. We start by
examining the overall algorithm performance and stability over
consecutive weeks. We then evaluate the optimal core size, i.e., a
core that results in a minimal inference mistake while achieving a
high classification percentage. Finally, we test the sensitivity of
the algorithm to errors in the core by randomly replacing core
vertices.

We start by looking at the result of executing the algorithm using
the first five weeks of 2007, each time with a different core type.
As expected, most of the AS relationships are inferred in phase 1 of
the deterministic algorithm using paths that traverse the core.
These paths comprise a large percentage of all available paths,
ranging from over 98\% for \kCore and \CP to 81\% for the smaller
\GMC core.

Table~\ref{table:core-type} shows the structure of the different
core types used and the effect it has on the deterministic
inference algorithm. The percent of classified edges and edges
matching CAIDA's inference is calculated out of the total number
of edges (including edges that are unclassified by CAIDA). It
shows that the smallest \GMC core results in the lowest
deterministic inference percentage while the largest \CP core have
the highest percentage. This is the result of the larger cores
having more paths that traverse them, therefore can be
deterministically inferred. \kCore provides an excellent overall
inference percentage (over 95\% deterministically inferred and
around 75\% matching CAIDA). Additionally, the results are stable
over the measured fix weeks period. The drop in the percentage of
edges matching CAIDA in week 5 is caused due to a decrease in the
number of edges classified by CAIDA.
\begin{table}[htb]
 \centering \scriptsize
  \caption{Structure of input cores and its effect on the deterministic inference algorithm} \label{table:core-type}
\begin{tabular}{|p{0.05\textwidth}|p{0.1\textwidth}|c|c|c|c|c|}
\hline
Core & \multicolumn{1}{r|}{Week$\rightarrow$} & 1 & 2 & 3 & 4 & 5\\
\hline
\kCore & Core Vertices & 57 & 56 & 54 & 58 & 54 \\
\cline{2-7}
 & Core Edges & 2260 & 2198 & 2076 & 2344 & 2134 \\
\cline{2-7}
 & Classified & 95.59\% & 95.76\% & 95.34\% & 95.24\% & 94.22\% \\
\cline{2-7}
 & Match CAIDA & 75.23\% & 75.32\% & 75.08\% & 73.76\% & 62.76\%  \\
\hline
Greedy & Core Vertices & 17 & 17 & 17 & 18 & 17 \\
\cline{2-7}
Max & Core Edges & 272 & 272 & 272 & 306 & 272 \\
\cline{2-7}
Clique & Classified & 89.64\% & 89.87\% & 89.77\% & 89.62\% & 88.87\% \\
\cline{2-7}
 & Match CAIDA & 73.73\% & 73.82\% & 73.60\% & 72.56\% & 61.68\% \\
\hline
CP & Core Vertices & 1067 & 1053 & 1068 & 1056 & 1087 \\
\cline{2-7}
& Core Edges & 6158 & 6110 & 6012 & 5844 & 6138 \\
\cline{2-7}
 & Classified & 98.29\% & 98.55\% & 98.45\% & 98.0\% & 97.39\% \\
\cline{2-7}
 & Match CAIDA & 79.77\% & 79.78\% & 79.43\% & 77.93\% & 67.19\% \\
\hline
\end{tabular}
\end{table}

Although \CP core seems to result in the best overall performance,
constructing the \CP core revealed that only a few p2p edges out of
the approximately 6,000 edges were not a part of the largest
connected component. This suggests that CAIDA incorrectly infers AS
relationships as p2p, since it is highly unlikely that almost all p2p edges
are connected. This causes a bias, resulting in more inference
errors.

\ignore{ This raises additional benefit in using combined DIMES and
BGP data. - DIMES uses almost 2,000 distributed measuring agents
every week, meaning that a moderate churn of agents is tolerable and
does not cause degradation in the quality of the data.}

Table~\ref{table:comapre-cores} shows that less than 6\% of the
edges were differently classified using two cores in each week, and
the difference between \kCore and \GMC is much smaller. This shows
that the algorithm results are relatively consistent regardless of
the input core.
\begin{table}[htb]
 \centering \scriptsize
  \caption{Percentage of edges that change classification comparing different core types}
 \label{table:comapre-cores}
\begin{tabular}{|p{0.15\textwidth}|c|c|c|c|c|}
\hline
Cores~~~~~~~~~~$|$~~~Week$\rightarrow$ & 1 & 2 & 3 & 4 & 5\\
\hline
\kCore~- \GMC & 1.77\% & 1.66\% & 1.58\% & 1.8\% & 1.64\% \\
\hline
\kCore~- \CP & 5.94\% & 5.89\% & 5.84\% & 5.7\% & 5.81\%  \\
\hline
\GMC~- \CP & 3.53\% & 3.45\% & 3.37\% & 3.34\% & 3.51\% \\
\hline
\end{tabular}
\end{table}

In order to find the best core size, we run the algorithm with a
growing core size starting at four vertices.  We do this for two of
our core types - \kCore and \GMC, using the first week of 2007. We
start with the highest degree vertices and add vertices in a
non-increasing degree order. Using \kCore, we first add vertices
from the $K_{max}-Core$ and then proceed to shells with lower
indexes.
\begin{figure*}[bt] \centering \scriptsize
\subfigure[\kCore]{
    \label{fig:kcore-coreSize}
    \includegraphics[width=.35\textwidth]{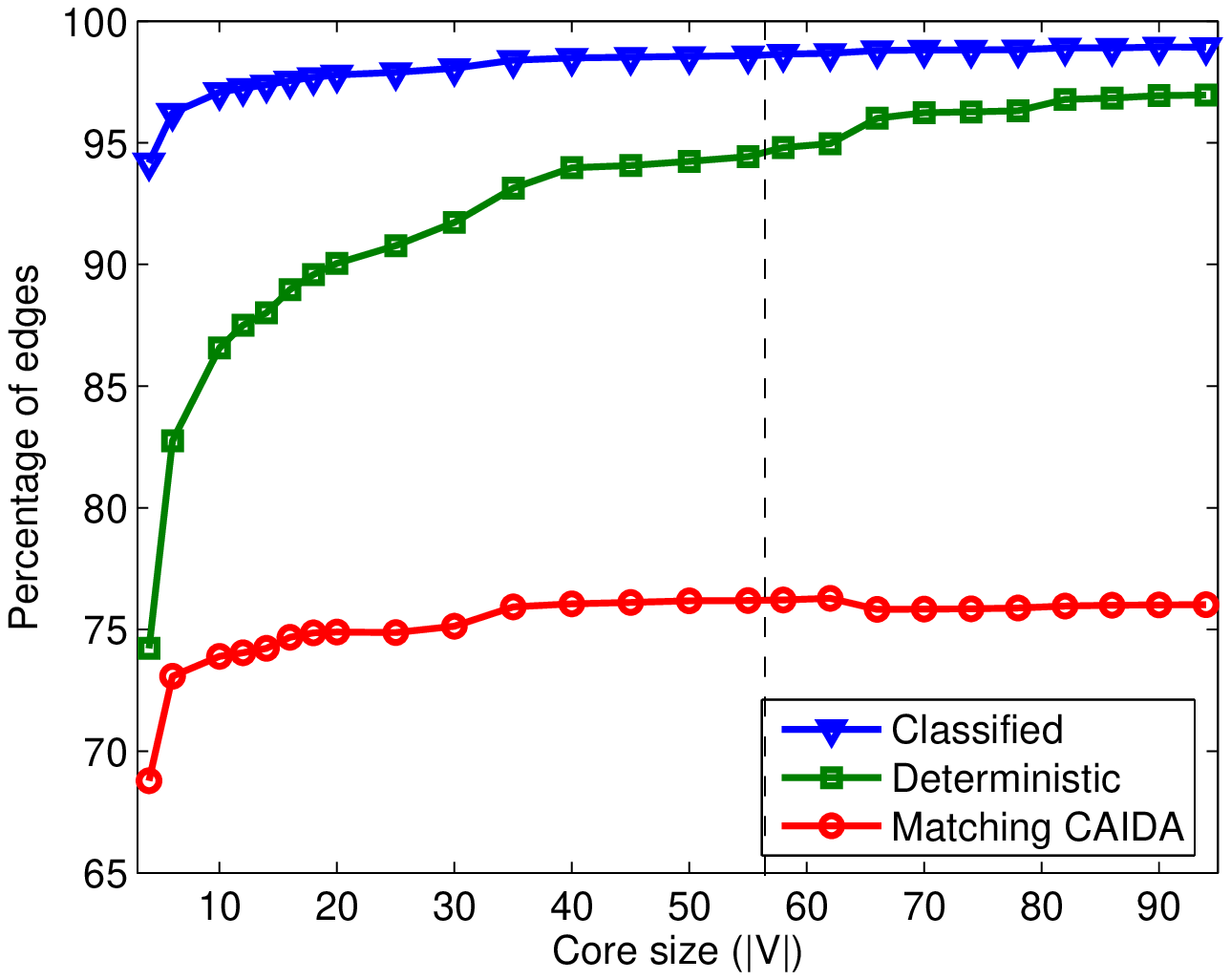}}
    \hspace{.3in}
    \subfigure[\GMC]{
    \label{fig:maxClique-coreSize}
    \includegraphics[width=.35\textwidth]{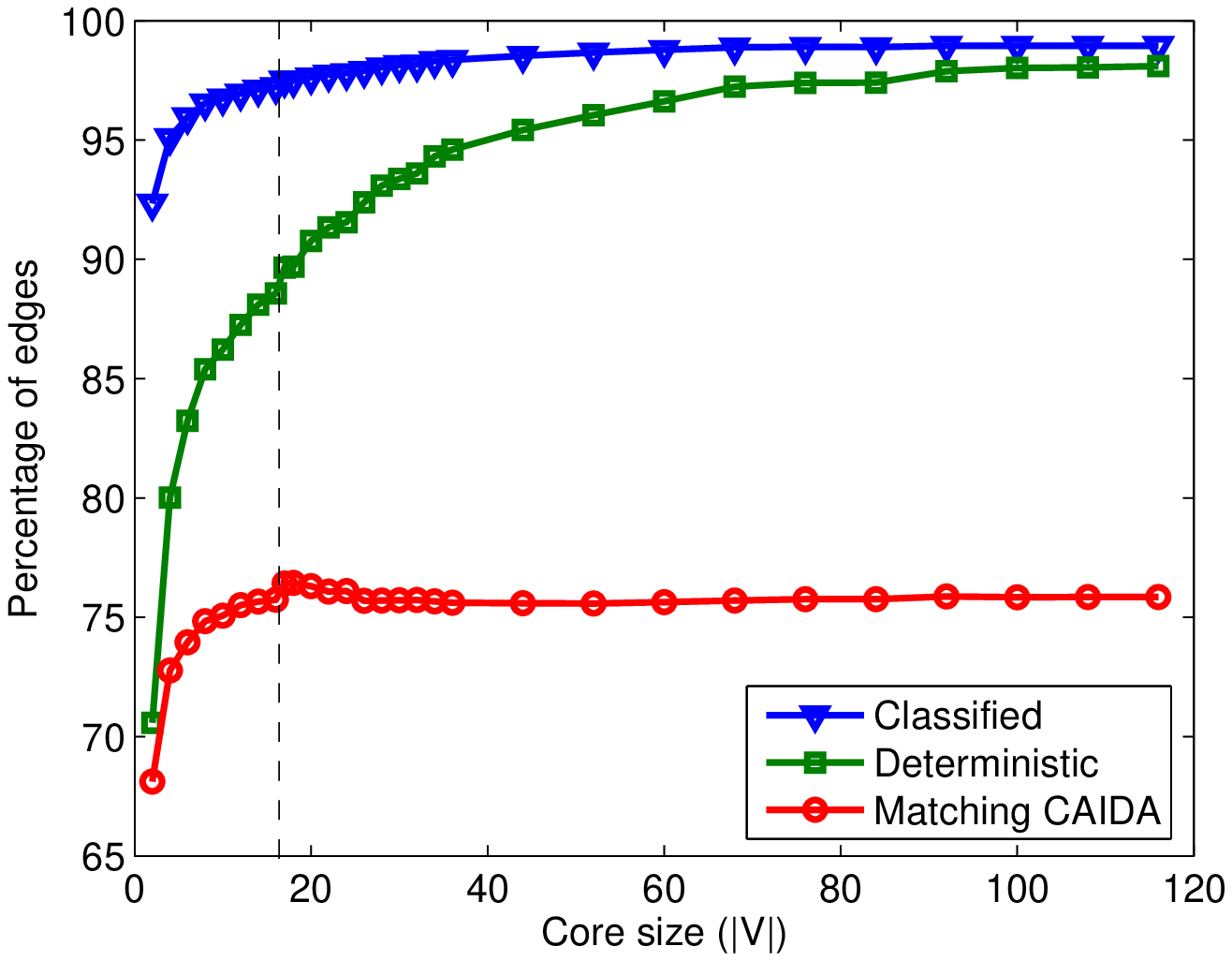}}
    \caption{Robustness of the algorithm to changes in the size of the core}
    \label{fig:robust-coreSize}
\end{figure*}

\begin{figure}[tbh]
\centering
\includegraphics[width=.35\textwidth]{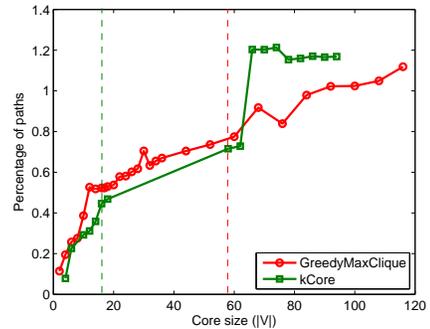}
\caption{Non-Valley-Free paths} \label{fig:non-valley-free-coreSize}
\end{figure}

\figref{fig:robust-coreSize} shows the robustness of the algorithm
relative to the size of the input core.  The vertical dashed line
marks the true core size.  For both cores, it shows that for more
than 20 vertices in the core the algorithm classification success
and similarity to CAIDA do not significantly change, while the
number of deterministically classified edges increases. However,
this increase comes with an increase in the percentage of
non-valley-free paths as shown in
\figref{fig:non-valley-free-coreSize}. This implies that the core
must be kept small enough to decrease the number of invalid paths.

Overall we showed that the algorithm is consistent over time and
various cores. Additionally, a core containing approximately 20
top-level ASes is sufficient to obtain excellent inference results.

\subsection{Time Aggregation Analysis}
\label{subsec:agg-periods}We wish to find a time frame for which the
algorithm captures best the relationships between ASes. A short time
frame results in a fast running algorithm but might miss AS links
and AS paths, especially in the DIMES data. This results in a low
vote count, possibly decreasing the success of the algorithm. On the
other hand, a long time frame captures two effects that can also
cause a decrease in the algorithm's success: 1) commercial
relationships can change and complex routing behaviors may occur
over long durations, and 2) possible measurements mistakes can pile
up and skew the results. \ignore{If the algorithm is viewed as an
attempt to take a snapshot of the Internet commercial relationships,
then using a too-short time period results in an image that is not
"sharp" enough, while using a too-long time period results in an
image that is "blurred".}

We executed the algorithm on an increasing time frame. We started
with the first day of 2007 and aggregated single days until the end
of the first week (for this daily analysis we used three RouteViews
files a day). Then, we aggregated a week at a time, until reaching
10 consecutive weeks. DIMES provides approximately 1.5M non-unique
tracroutes each day, reaching over 100M traceroutes for the 10 weeks
period. RouteViews provides approximately 1.2M \emph{unique} paths
regardless of the time frame used.
\begin{figure}[htb]
\centering
\includegraphics[width=.35\textwidth]{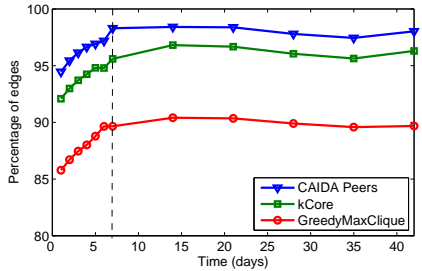}
\caption{Deterministically classified edges over an increasing time
frame} \label{fig:deterministic-period}
\end{figure}

The percentage of deterministically inferred edges over the
aggregated time frame is shown in \figref{fig:deterministic-period}.
Using data from a single week (marked as the vertical dashed line)
results in over 90\% of the edges being classified for all core
types, having \CP obtaining the best percentage and \GMC the worst.
This is directly related to the size of the core, since a larger
core results in more paths that traverse through it, yielding more
deterministically inferred relationships. \ignore{The percentage
remains fairly stable with a very small decent for longer time
frame. However, the results for the longest time frame we measured
(10 weeks) are even better than when using a smaller time frame.
This is probably due to short-term routing changes that
statistically "disappear" as the more common routes become dominant
over time.}

\begin{figure}[htb]
\centering
\includegraphics[width=.35\textwidth]{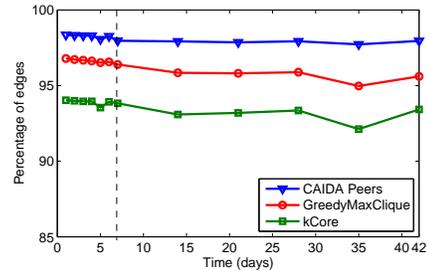}
\caption{Percentage of deterministically inferred edges matching
CAIDA out of edges that are inferred by both algorithms}
\label{fig:deterministic-match-caida-period}
\end{figure}

We evaluate the deterministic algorithm over this time frame by
looking at the edges that are identically classified by the
deterministic algorithm and CAIDA, out of the edges that are
classified by both. \figref{fig:deterministic-match-caida-period}
shows that for all cores and any time frame, the algorithms agree on
over 92\% of the edges. Obviously, using \CP gives the best match to
CAIDA's inference. It is interesting to see that although \kCore has
better overall deterministic inference success than \GMC (shown in
\figref{fig:deterministic-period}), it results in a lower match
rate. This is probably due to the relatively small, local and
degree-based \GMC core, which is more related to the heuristic used
by CAIDA's inference methods.

Finally, we looked at the consistency of the inference results over
the time frame by comparing edges that are classified in both time
frames. We found that over 98\% of the inferences remain constant
between consecutive time frames. This suggests that there are only a
few commercial relationships that change over time. Short-term
routing changes have very little effect, since they statistically
"disappear" as the more common routes become dominant over time.

\subsection{Analysis of Non-Deterministically Inferred Relationships}
\label{subsec:exp-unclassified-edges} Edges that the deterministic
algorithm fails to classify are classified using the two
heuristic-based inference methods described is
section~\ref{sec:unclassified-algorithm}. The first method breaks
voting ties and the second infers p2p relationships.

To break voting ties we compared adjacent AS degrees (similar
to~\cite{gao01}) and inferred the relationship between them as p2p
if the degrees ratio is between 0.8 and 1.2, or p2c otherwise
(marking the provider as the AS with the higher degree). For \kCore
we also compared the $k$-Shell index, and inferred the relationship
to be p2p if the two ASes have the same $k$-Shell index, or p2c
otherwise (marking the provider as the AS with the higher $k$-Shell
index) and note very little difference between the two heuristics.

\begin{figure*}[tbh]
\centering \subfigure[\kCore]{
    \label{fig:kCore_random}
    \includegraphics[width=.35\textwidth]{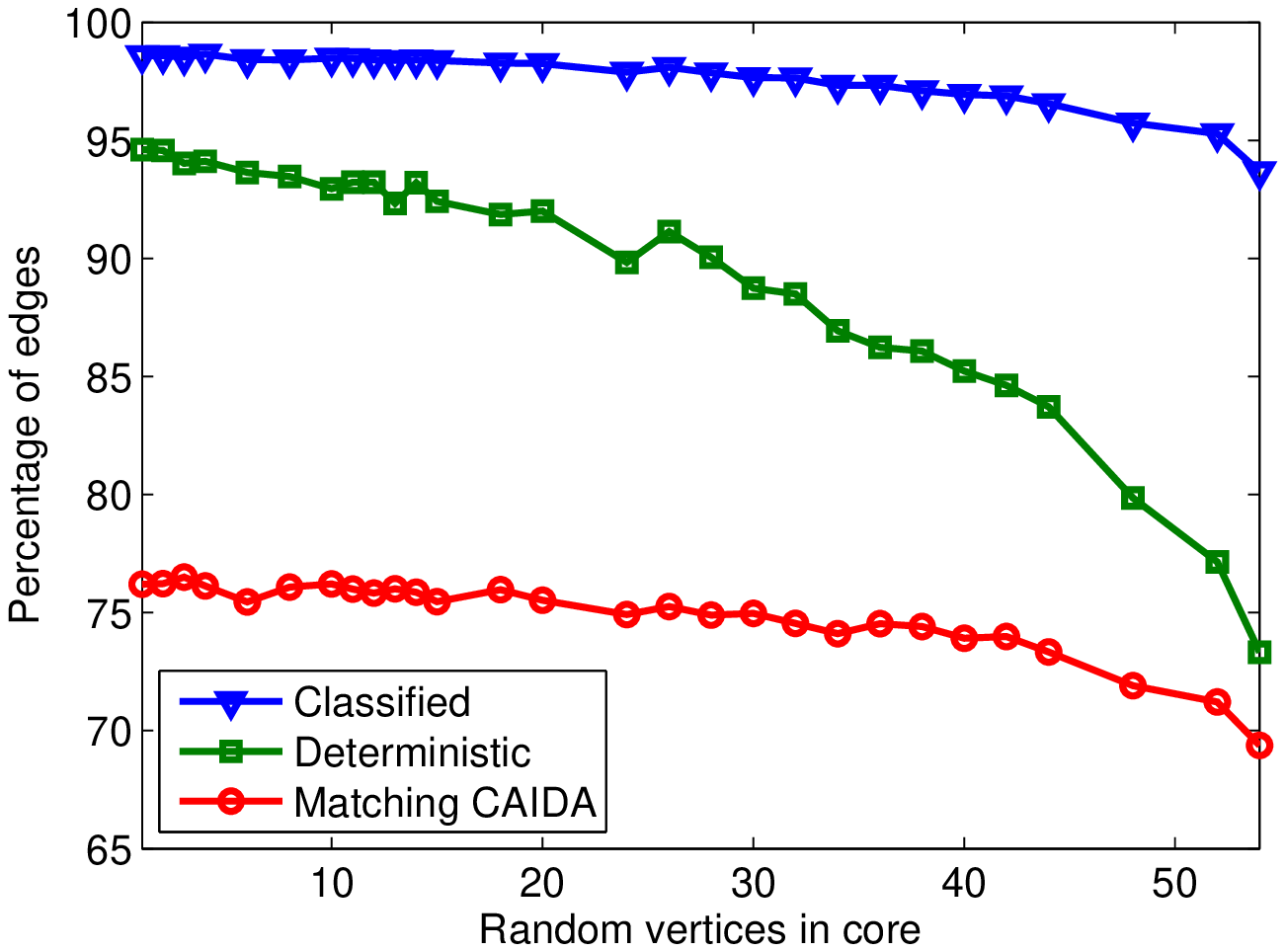}}
    \hspace{.3in}
    \subfigure[\GMC]{
    \label{fig:maxClique_random}
    \includegraphics[width=.35\textwidth]{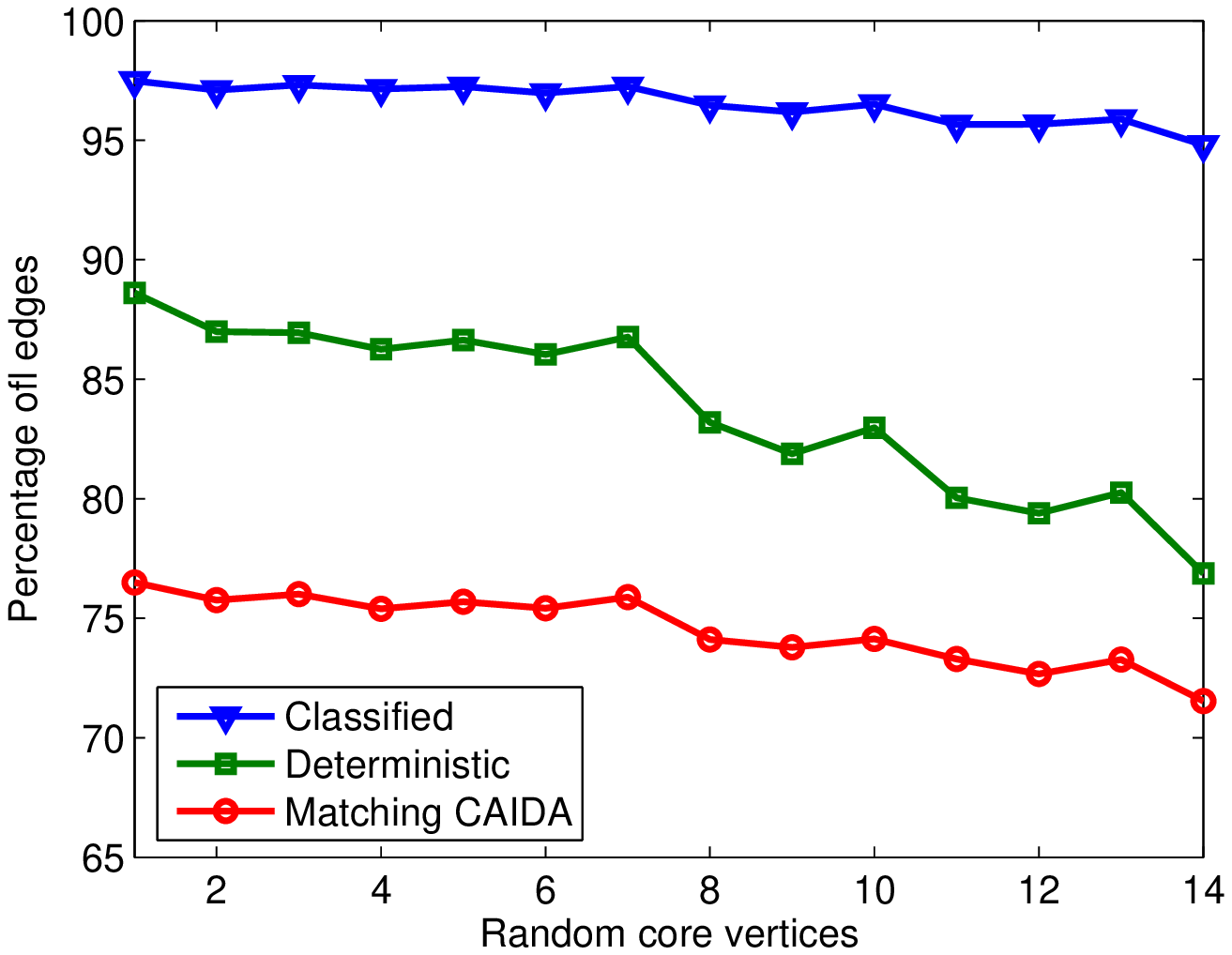}}
    \caption{Robustness of the algorithm to errors in the core}
    \label{fig:random}
\end{figure*}

We estimate the accuracy and robustness of the algorithm by
intentionally increasing the mistake in the core. We do this by
randomly replacing ASes in the core and see how the number of
relationships inferred and their correlation with CAIDA's inference
change. We start by replacing one core AS with one random AS (that
is connected to at least one of the remaining core ASes) and
gradually replace more ASes until we have a core that consists of
completely random but still connected ASes. \figref{fig:random}
shows the percentage of classified edges, deterministically
classified edges and the percentage of edges that match CAIDA's
inference using \kCore and \GMC.

\begin{figure*}[tbh]
\centering \subfigure[\kCore]{
    \label{fig:kcore-heu-random}
    \includegraphics[width=.35\textwidth]{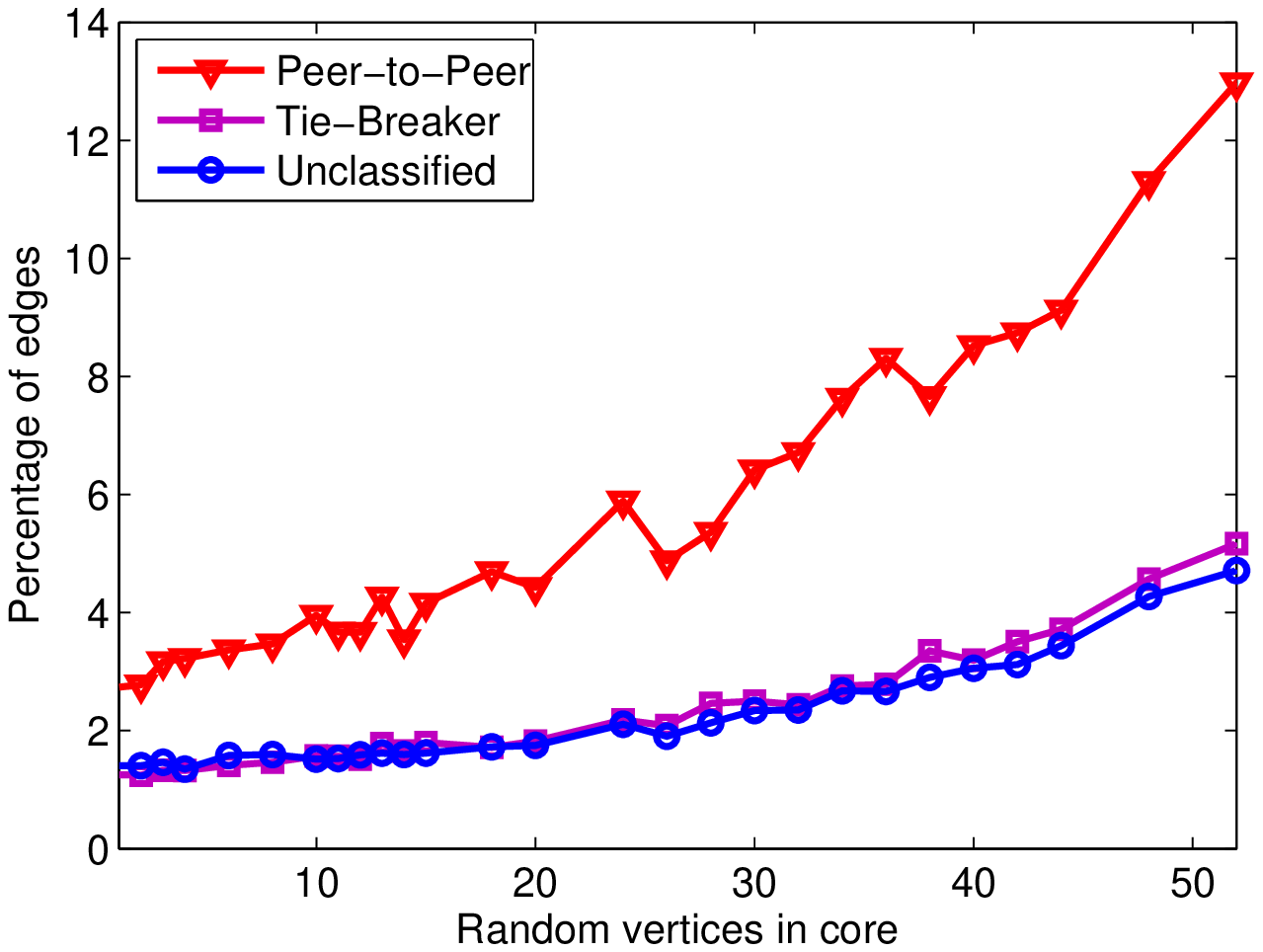}}
    \hspace{.3in}
    \subfigure[\GMC]{
    \label{fig:maxClique-heu-random}
    \includegraphics[width=.35\textwidth]{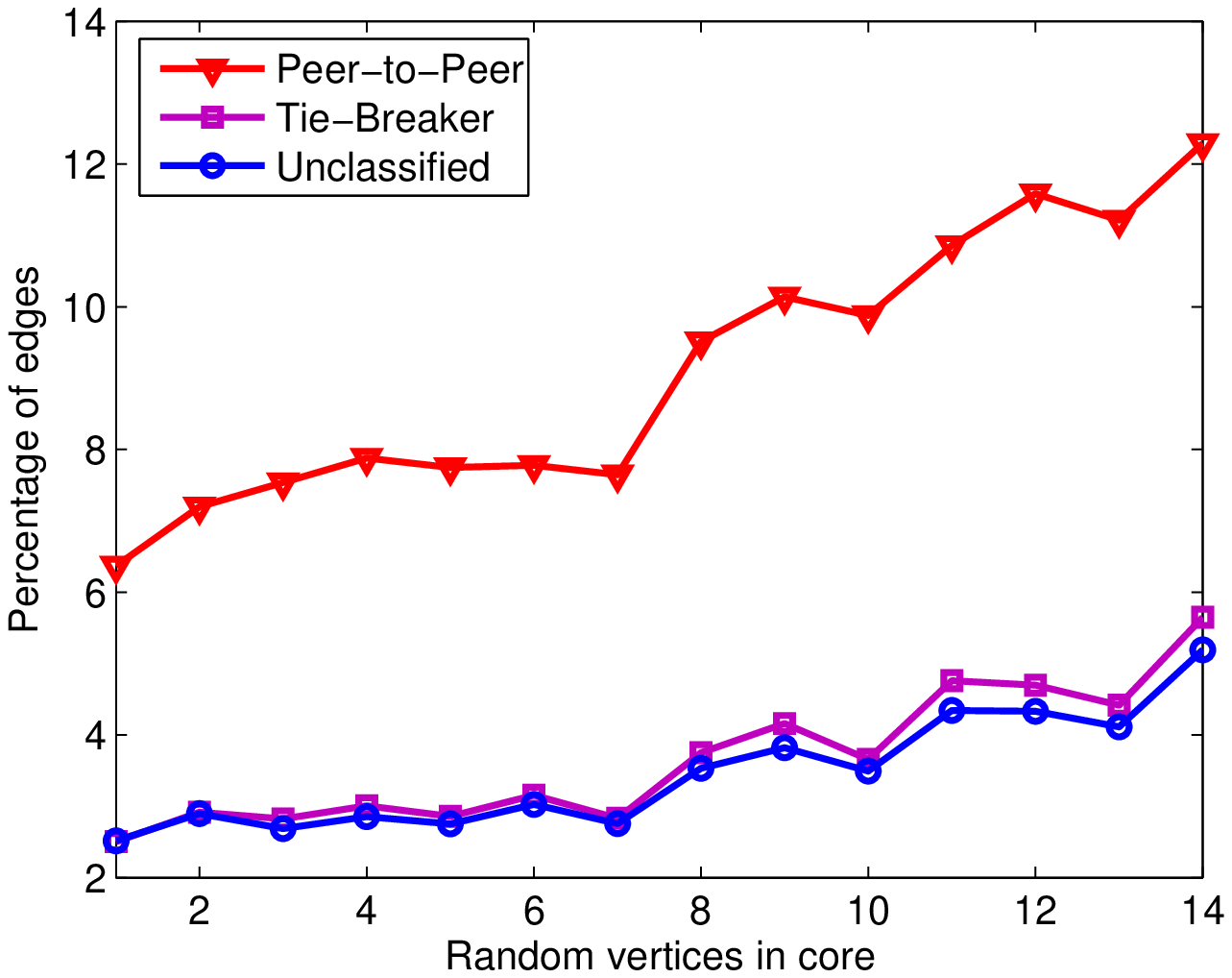}}
    \caption{Percentage of p2p, heuristically classified and unclassified edges}
    \label{fig:core-heu-random}
\end{figure*}

Interestingly, while the algorithm's performance decreases as we
increase the randomness of the core, the overall degradation is not
as high as one would expect. \ignore{ We can see that \GMC results
in less than 8\% heuristically inferred edges out of total number of
edges, while \kCore heuristically inferred less than 3\% of the
edges. This is mainly the result of \kCore being much larger than
\GMC. Analyzing the robustness of the algorithm to increasing
mistake in the core, we can see that as random vertices are inserted
to the core the number of classified edges decrease, having the
heuristics method compensate the deterministic algorithm that
suffers due to increasing number of non-valley-free paths. However,
the number of edges that match CAIDA's classification decrease only
in 5\% (out of the total number of edges), even for a completely
randomly selected core. }
However, \figref{fig:core-heu-random} shows a rising trend of the
percentage of unclassified, p2p and tie-breaking heuristic edges as
we inject errors to the core. As more errors are injected, the
algorithm needs to use more heuristics. Particulary, when there are
approximately 50\% random vertices in the core, the effect of the
increasing mistake becomes more noticeable. However, even with a
completely random core, the overall heuristically inferred edges
account for less than 20\% of all edges.

These results indicate that although the algorithm seems quite
robust to the mistake in the core, it is still significantly
affected once there are more than 50\% incorrect core vertices.
Looking at p2p edges showed us that as we inject more errors, the
percentage of p2p edges that are classified differently by CAIDA
increases from around 16\% to over 40\%. \ignore{ Around 40\% of the
p2p edges are not classified by CAIDA, since they do not appear in
the BGP routing tables.
\begin{figure}[htb]
\centering
\includegraphics[width=.35\textwidth]{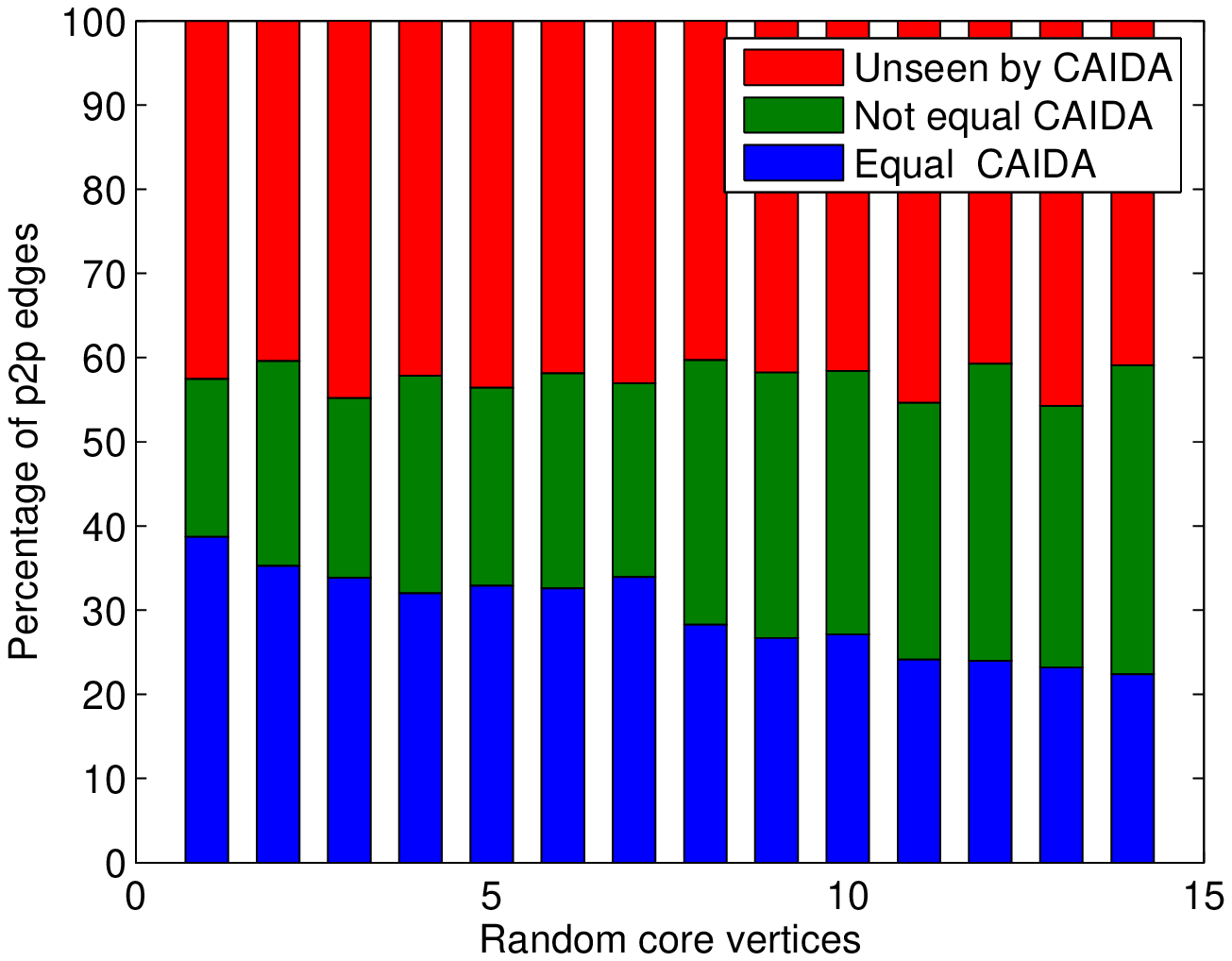}
\caption{P2P inference method using GreedyMaxClique core}
\label{fig:maxClique_p2p_random_bar}
\end{figure}
}

Finally, we analyzed the distribution of relationship types for
edges that appear only in DIMES, and are not seen in the BGP routing
tables, to understand what are the types of relationships that DIMES
discovers. We found that while on average the p2p relationships
comprise 4-5\% of the total number of edges, it goes up to around
12\% of the edges that appear only in DIMES. Moreover, approximately
40\% of the p2p edges inferred by our algorithm, do not appear in
the RV tables. This means that utilizing DIMES significantly
improves the ability to detect p2p links between ASes, mainly since
DIMES agents are spread over the Internet and contribute AS links
that are either not collected by the RouteViews routers or even not
published in the BGP protocol.

\section{Conclusion and Future Work}\label{sec:conclusion}
The common weakness of previously proposed AS relationships
inference algorithms is their lack of guarantee on inference errors
introduced during the process. This work improves on existing
methods by providing a near-deterministic algorithm that, given a
classified error-free input core, does not introduce additional
inference errors. We investigate various input cores and show that
the proposed algorithm provides accurate inferences that are robust
under changes in the core's size and creation technique. We show
that a core containing as little as 20 almost fully-connected ASes
is sufficient for good inference results. Additionally, we show that
heuristic methods can still play an important role in inferring the
remaining relationships. Using data collected from a single week
(containing approximately 1.2M BGP paths and over 10M DIMES AS-level
traceroutes), the algorithm runs for only about 2 hours and yields
over 95\% deterministically inferred relationships.

As the Internet grows larger, many providers interconnect at
multiple locations for traffic engineering and embrace the usage
of exchange points. The relationships and policies used in these
interconnection points might not conform to either
provider-to-customer or peer-to-peer relationships. Moreover, it
might not even conform to the valley-free property. The data
provided from the DIMES project can reveal these complex
relationships and seed other large scale Internet analysis work.

\bibliographystyle{abbrv}
\bibliography{classification}

\end{document}